\documentclass[useAMS,usenatbib,usegraphicx]{mn2e}

\usepackage{amssymb}
\usepackage{natbib}

\title[Thin discs and thick dwarfs]{Thin discs, thick dwarfs and the effects of stellar feedback}

\author[S\'anchez-Janssen et al.]{R. S\'anchez-Janssen $^{1}$\thanks{E-mail:
rsanchez@eso.org}, J. M\'endez-Abreu$^{2,3}$ and J.A.L. Aguerri$^{2,3}$\\
$^{1}$European Southern Observatory, Alonso de C\'ordova 3107, Vitacura, Santiago, Chile\\
$^{2}$Instituto de Astrof\' isica de Canarias, Calle V\'ia L\'actea s/n, E-38200 La Laguna, Tenerife, Spain\\
$^{3}$Departamento de Astrof\' isica, Universidad de La Laguna, E-38205 La Laguna, Tenerife, Spain}

\begin{document}

\date{2010 May 21}

\pagerange{\pageref{firstpage}--\pageref{lastpage}} \pubyear{2010}

\maketitle

\label{firstpage}

\begin{abstract}
We investigate the role of stellar mass in shaping the intrinsic thickness of galaxy discs by determining the probability distribution of apparent axis ratios ($b/a$) for two different samples that probe the faint end of the galaxy luminosity function. We find that the $b/a$ distribution has a characteristic 'U-shape' and identify a limiting mass M$_{*} \approx 2\times10^{9}$ M$_{\odot}$ below which low-mass galaxies start to be systematically thicker. This tendency holds for very faint  ($M_{B} \sim -8$) dwarfs in the Local Volume, which are essentially spheroidal systems. We argue that galaxy shape is the result of the complex interplay between mass, specific angular momentum and stellar feedback effects. Thus, the increasing importance of turbulent motions in lower mass galaxies leads to the formation of thicker systems, a result supported by the latest hydrodynamical simulations of dwarf galaxy formation and other theoretical expectations. We discuss several implications of this finding, including the formation of bars in faint galaxies, the deprojection of H{\sc i} line profiles and simulations of environmental effects on the dwarf galaxy population. 
\end{abstract}


\section{Introduction}
Galaxy discs are key structural components in the understanding of galaxy formation. They contain approximately 60\% of the stellar mass in the Universe \citep{Driver2007} and are one of the main sites of current star formation activity \citep{Kennicutt1998}. Moreover, their prominence and appearance form the basis of the \citet{Hubble1926} morphological sequence.

Despite this enormous importance, understanding the details of disc formation remains extremely challenging. The basic scenario considers that baryons cool and collapse within hierarchically assembled dark matter haloes. A certain amount of their torque-acquired angular momentum is transferred to the baryonic component leading to the formation of a rotationally supported thin disc \citep{Fall1980}. This relatively simple picture has been able to reproduce many observational properties of disc galaxies --flat rotation curves, Tully-Fisher relation, gas content (e.g., \citealt{Dalcanton1997,Mo1998,vandenBosch1998,vandenBosch2000})-- but suffers from several shortcomings. On the one hand, detailed N-body simulations showed that when the dissipative effects of gas are not considered, the hierarchical nature of structure formation unavoidably results in disc destruction \citep{Toth1992}. On the other hand, hydrodynamical simulations of this process systematically produced discs that were too small and too centrally concentrated due to excessive angular momentum exchange between the gas and the dark matter haloes \citep{Navarro1997}. 

A physically-motivated solution for these problems is the inclusion of strong feedback effects from different sources --but specially from star formation and supernovae explosions-- which, when coupled with a cosmic UV field, are able not only to produce realistic discs (e.g., \citealt{Governato2010}), but also to provide one possible explanation for the missing satellite problem \citep{Klypin1999}.

This approach, in turn, raises the question of which role does mass play in shaping the properties of discs, as these strong heating mechanisms are expected to produce a greater influence in lower-mass galaxies (\citealt{Kaufmann2007}, KWB07 hereafter). Indeed, it is well known that galaxies have higher gas mass fractions (e.g., \citealt{Schombert2001}) and more extended star formation time scales \citep{Hunter1985} as they are less massive, and it has been proposed that dwarf and disc galaxies are probably two different structural entities \citep{Schombert2006}.

From an observational point of view, galaxy discs are best described as flattened triaxial ellipsoids with exponential surface brightness profiles \citep{Freeman1970}. Though they are traditionally considered to be perfectly circular,  it is well known that discs are indeed slightly elliptical, with $b/a \gtrsim 0.9$ \citep{Lambas1992,Ryden2004}. They are said to be \emph{thin} because their vertical to radial axis ratios have long been known to lie in a narrow range $0.15 \lesssim q_{0} \lesssim 0.25$ \citep{Holmberg1950,Sandage1970,Heidmann1972}. However, this thickening is not uniform, as late-type spirals have thinner discs than early-types \citep{Bottinelli1983,Guthrie1992}. If we neglect these small ellipticity deviations, discs can then be thought of as axisymmetric oblate spheroids with intrinsic thickness $q_{0}$. In that case, the apparent axis ratio of a disc is determined by $b/a = \sqrt{q_{0}^{2}+(1-q_{0})^{2}\cos^{2}i}$, where $i$ is the corresponding inclination angle. Interestingly, the apparent axis ratio distribution of a randomly oriented population of such discs peaks at a minimum value of $q_{0}$ --i.e., edge-on galaxies are much more common than any other projection. This property implies that the intrinsic thickness of a disc population can be identified as the minimum value of the distribution of apparent axis ratios.

In this paper we take advantage of this prediction to investigate the range of masses where thin discs exist. For this purpose we study the probability distribution of $b/a$ as a function of galaxy mass (luminosity) for two different samples that probe the faint end of the galaxy luminosity function. In Section\,2 we develop further on the samples characteristics and the methodology. Section\,3 presents the main results, which are discussed in Section\,4 together with additional implications of this investigation. Throughout this Letter we adopt a $\Lambda$CDM cosmology with $\Omega_{0} = 0.3$, $\Omega_{\Lambda} = 0.7$ and $h_{75} = H_{0}/(75$ km\,s$^{-1}$\,Mpc$^{-1})$.

\section[]{Sample and methodology}
Two different galaxy samples have been used throughout this work. 
The first sample consists of \emph{all} 9245 galaxies in the SDSS-DR7 \citep{Abazajian2009} with recession velocities in the $2000 < cz < 6000$ km\,s$^{-1}$ range. The lower limit was set to avoid strong corrections in galaxy distances due to the Virgo infall velocity field \citep{Tammann1985}. The upper one ensures that the sample is volume-limited for galaxies brighter than $M_{i} \approx -16.5$, additionally including 2063 fainter objects down to a limiting magnitude of $M_{i} \approx -14.5$. 
We however note that this incompleteness only affects the statistical significance, but not the shape, of the axis ratio probability distribution at the lowest luminosities, as the volume we probe is too narrow ($25 \lesssim D \lesssim 80$ Mpc) for any shape evolution to take place.
A further concern might arise given the well-known incompleteness of the SDSS imaging for low surface brightness galaxies (LSB; \citealt{Blanton2005}). However, if a population of LSB discs exists, we will preferentially miss the face-on (high $b/a$) systems instead of the edge-on, higher surface brightness discs. Therefore, the fraction of highly flattened galaxies is in any case a lower limit.

The SDSS imaging provides $ugriz$ magnitudes and several estimates of axis ratios in each band. For this study we have used the 25 mag\,arcsec$^{-2}$ isophotal axis ratios, which provide a robust measurement of galaxy shapes at their outer regions --a few times the galaxy's effective radius \citep{Vincent2005}--, where the effects of dust and the presence of bulges are less important. 
Seeing effects do not play a role either, as the proximity of the sample ensures that essentially all galaxies have minor axis sizes larger than the typical seeing FWHM, with a median of 20 arcsec in the $r$-band. 
SDSS isophotal $b/a$ can sometimes take unrealistic values in a given photometric band. To avoid these, we have used the median of the three high signal-to-noise $gri$ axis ratio estimates as our reference $b/a$ value, therefore producing a less noisy ($\sigma_{b/a} \approx 0.05$) shape estimate for each galaxy. We however verified that our results are independent of the selected filter.


Given that the main purpose of this work is to study the influence of galaxy mass in the formation of discs, we have used the \citet{Bell2003} M/L ratios with a \citet{Kroupa2001} initial mass function in the transformation from luminosities to stellar masses.


In order to extend our analysis to even fainter luminosities, we also analyse the catalogue of neighbouring galaxies of \citet{Karachentsev2004}. This is an all-sky catalogue including 445 galaxies brighter than $M_{B} \lesssim -8$ with individual distance estimates $D \leq 10$ Mpc and is expected to be $70\%$-$80\%$ complete within 8 Mpc. From this catalogue we used the $B$-band absolute magnitudes and the 25 mag\,arcsec$^{-2}$ isophotal apparent axis ratios $b/a$ of each individual galaxy.

\section{Results}

\begin{figure*}
\includegraphics[width=0.33\textwidth,clip=true]{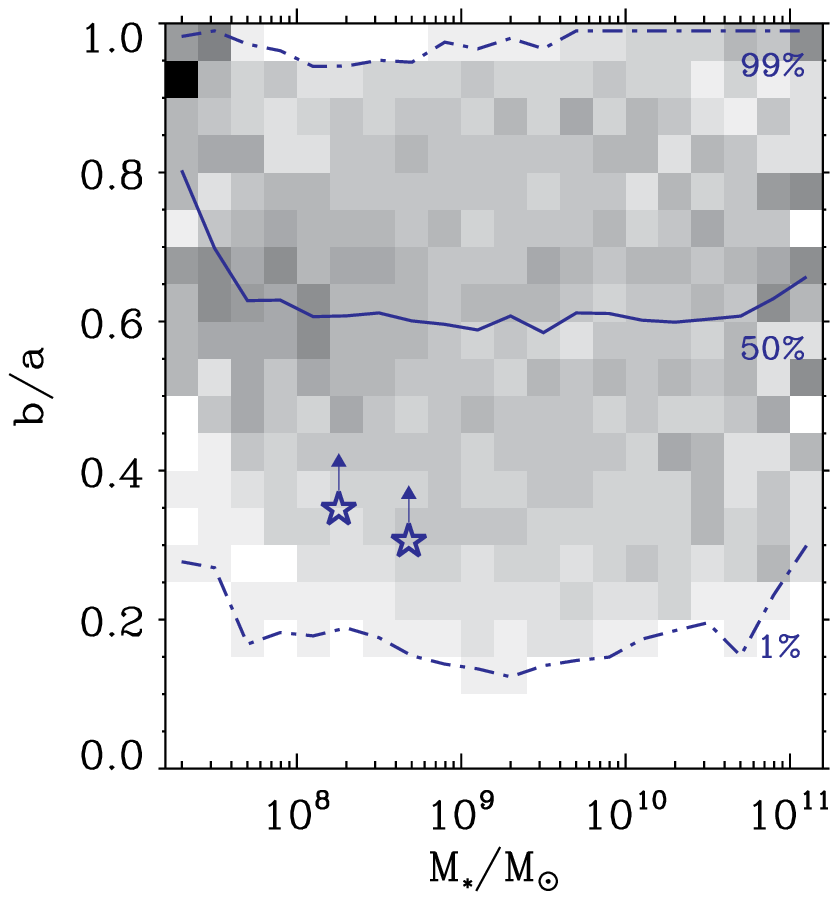}
\includegraphics[width=0.33\textwidth,clip=true]{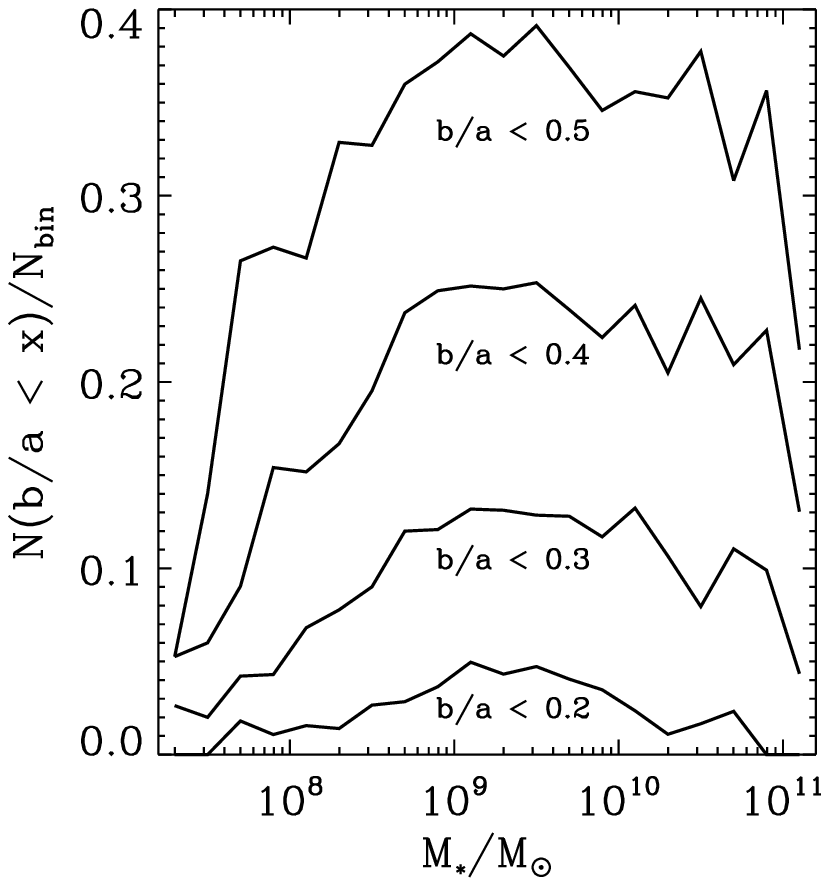}
\includegraphics[width=0.33\textwidth,clip=true]{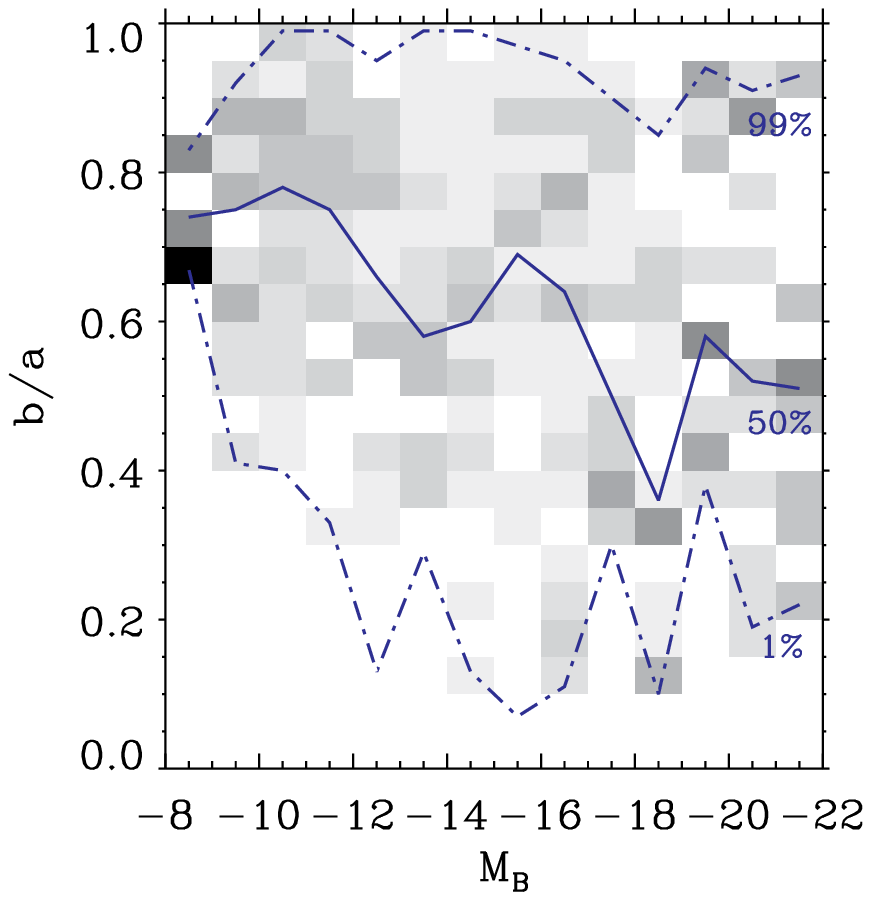}
\caption{
\emph{Left}: the grey scale represents the probability distribution of apparent axis ratios in intervals of stellar mass for the SDSS sample --where dark (light) indicates a high (low) probability within each M$_{*}$ bin. The median (solid line) and the 1\% and 99\% percentiles (dot-dashed lines) are also indicated. The distribution has a characteristic 'U-shape', showing a trend of increasing axis ratios for lower (and higher) mass galaxies. Recall that the typical $b/a$ uncertainty is $\approx$0.05, i.e., the binsize. The starred symbols indicate the edge-on thickness of the dwarf galaxies simulated by \citet{Governato2010} --and are therefore lower limits for their observed $b/a$.
\emph{Middle}: each curve shows the fraction of galaxies in each mass interval having a lower $b/a$ than the specified value.  The peak of the distribution occurs at a characteristic mass M$_{*} \approx 2\times10^{9}$ M$_{\odot}$, which we identify as the minimum stellar mass of thin disc galaxies.
\emph{Right}: same as in the left panel but for the Local Volume sample, that includes galaxies down to $M_{B} = -8$. Note the lack of faint flattened systems.
}
\label{fig:1}
\end{figure*}

In the left panel of Fig.\,\ref{fig:1} we present the probability distribution (grey scale) of apparent axis ratios in intervals of stellar mass for the SDSS sample. This figure contains several interesting features.

First, the fact that, aside from the most massive bins, the $99\%$ percentile (upper dot-dashed line) slightly deviates from unity, consistent with the small ellipticity typical of discs (e.g., \citealt{Ryden2004}).
Second, and most important for our purposes, both the median and the $1\%$ percentile (solid and lower dot-dashed line, respectively) of the $b/a$ distribution present a characteristic 'U-shape', with a minimum $b/a \approx 0.15$ located at M$_{*} \approx 2\times10^{9}$ M$_{\odot}$\,\footnote{Recall that, as previously pointed out, the lower percentile can be interpreted as the intrinsic thickness of the flattest galaxy population in a given stellar mass interval.}. We note that this value corresponds to $M_{i} \sim -18$, a much brighter luminosity than the limiting magnitude of our volume-limited sample. The minimum in the $b/a$ distribution is better appreciated in Fig.\,\ref{fig:1} (middle), where we plot the fraction of galaxies having an axis ratio smaller than a given value as a function of stellar mass. The distribution is indeed a strong function of M$_{*}$, and the fraction of low-mass thin galaxies never exceeds a few per cent. 

The increase of apparent axis ratio in the high mass range can be ascribed to two effects.  The main reason is that the contribution of the  spheroidal component is more important at higher M$_{*}$, being totally dominant for the brightest ellipticals (cf. \citealt{vanderWel2009}). Additionally, early-type (more massive) spirals tend to have thicker discs than their late-type counterparts \citep{Bottinelli1983}.

The increasing thickness towards lower masses is perhaps more puzzling given the previous fact and the decreasing bulge mass fraction at faint luminosities. Galaxies with stellar masses M$_{*} \sim 10^{7}$ M$_{\odot}$ are most probably spheroidal systems, with apparent axis ratios in the $0.3 \lesssim b/a \lesssim 0.95$ range and a median $b/a \sim 0.8$.
This paucity of thin discs in the dwarf regime is enhanced at the faintest magnitudes probed by the Local Volume catalogue (Fig.\,\ref{fig:1}, right), with less than $1\%$ of the $M_{B} > -14$ galaxies having $b/a < 0.3$. 


\section{Discussion and conclusions}
Hints of increasing thickness for less massive galaxies are not new (e.g., \citealt{Heidmann1972,Yoachim2006}), but to our best knowledge we are the first to identify a characteristic mass below which low-mass galaxies start to be systematically thicker. The question then arises naturally. What causes this effect? 

\begin{figure}
\centering
\includegraphics[width=0.4\textwidth,clip=true]{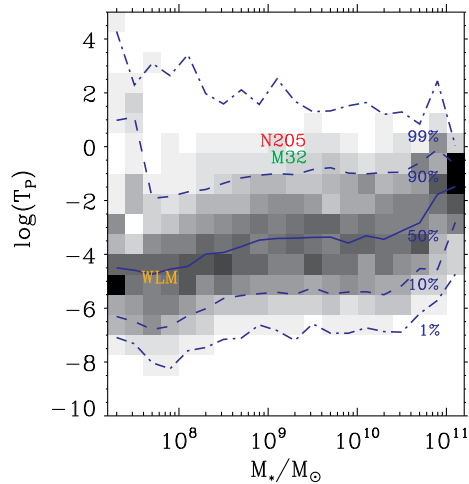}
\caption{
Probability distribution of the tidal parameter in bins of stellar mass (grey scale) for the SDSS sample. Note that low-mass galaxies are in general less tidally affected than their more massive counterparts. For reference we show the corresponding values for three local-volume dwarf galaxies experiencing different levels of interaction with M31.
}
\label{fig:2}
\end{figure}

One possibility is that environmental effects could influence galaxy shape. While the narrow volume we probe does not include any significant massive cluster --it is just further than Virgo and closer than Coma--, there remains the possibility that low-mass galaxies are tidally affected by more massive companions in group environments, as has been previously observed in the Local Group (\citealt{Choi2002}; \citealt{Geha2006a}). 
To test for this scenario, for each SDSS galaxy we have computed a \emph{tidal parameter} (e.g., \citealt{Varela2004}) $\mbox{T$_{P}$} = \mbox{max[(M$_{*,P}$/M$_{*,i}$)\,(A$_{25}$/D$_{P}$)$^{3}$}]$, where M$_{*,i}$ and  A$_{25}$ are its stellar mass and isophotal major axis, and  M$_{*,P}$ and D$_{P}$ are the corresponding stellar mass and projected distance to any potential perturber having a relative velocity $|\Delta\mbox{v}| < 1000$ km\,s$^{-1}$ and a luminosity $M_{i} < -16.5$. This parameter is proportional to the maximum ratio of external to internal forces that act on a galaxy, and takes into account that the effects of a closer, less massive companion can be even more important than those originated by a massive, more distant galaxy. 
Fig.\,\ref{fig:2} shows in grey scale the probability distribution of T$_{P}$ in each mass interval for the SDSS sample, while the different lines indicate the corresponding 1\%, 10\%, 50\%, 90\% and 99\% percentiles. There is a clear trend of decreasing T$_{P}$ with decreasing mass, i.e., low-mass galaxies are in general \emph{less} tidally affected than their more massive counterparts. This result is consistent with the fact that the galaxy luminosity function is dominated \emph{at all luminosities} by central galaxies \citep{Cooray2005}. Given that the normalisation of the tidal parameter is somewhat arbitrary, we have used the Hyperleda database \citep{Paturel2003} to compute the  T$_{P}$ of three dwarf galaxies experiencing different levels of interaction with M31 (see \citealt{Karachentsev2004}). We find that the fraction of strongly tidally perturbed galaxies like M32 or N205 is only $\sim$10\% , while the vast majority of low-mass galaxies are rather isolated systems as WLM. 

We therefore conclude that tidal effects are not responsible of the increasing thickness found in fainter galaxies, and suggest that the effect is probably related to the increasing importance of feedback mechanisms in low-mass haloes. 

KWB07 investigated the effect that a gas temperature floor ($T_{F}$) --as might arise in the presence of a cosmic UV field and/or due to stellar feedback-- has on the properties of low-mass galaxies. Their instructive fig.\,1 shows that the final morphology of a galaxy is the result of a complex interplay between the turbulent support provided by this $T_{F}$ and the angular momentum support, characterised by the dimensionless spin parameter $\lambda$. For low-mass haloes, the pressure support radius becomes comparable to or larger than the rotational support radius, and thus galaxies are naturally formed thicker.

\citet{Schombert2006} was the first to note a separation between dwarf and disc galaxies in the scale-length versus mass plane, in the sense that the former are more diffuse (extended) than the latter. He suggests that the increasing importance of turbulent motions in dwarf galaxies  results in thicker systems, potentially giving rise to the two observed sequences. However, he also points out that 'this scenario provides no explanation for why there are two distinct sequences rather than a smooth transition in structure from an oblate shape to a triaxial one as one progresses to lower galaxy mass'. We believe that the key to this separation resides again in the complex interplay between galaxy mass, heating mechanisms and specific angular momentum. 

Going back to KWB07's fig.\,1, we can identify three different regions. 
For the most massive haloes, angular momentum support prevails regardless of spin parameter or temperature floor value. In this region, thin discs will form, lying in Schombert's sequence defined by Sc-Sd galaxies.
In the case of very low-mass haloes, the opposite is true: pressure support is always dominant even for haloes with the highest spins $\lambda \sim 0.1$. Galaxies in this region will be puffier, creating the Im and dIrr sequence in the scale-length vs. mass relation.
Finally, for the intermediate mass region KWB07 point out that a range of morphologies is predicted if galaxies have spins within the expected range ($0.01 \lesssim \lambda \lesssim 0.1$; \citealt{Bullock2001}) and the temperature floor does not vary too much among galaxies. This would naturally explain the co-existence of the two sequences in the narrow $10^{9}$-$10^{10}$ M$_{\odot}$ mass range: galaxies in the high-spin tail of the distribution would end up in the disc sequence, while the dwarf sequence would be populated by galaxies with low spin parameter.

Evidences for the decreasing importance of ordered motions in favour of turbulent ones come from the high thickness of the neutral gas component in very faint galaxies from the FIGGS sample \citep{Roychowdhury2010}, or from the shape of H{\sc i} line profiles from the dwarf sample of \citet{Geha2006}, where only $18\%$ of the objects exhibit double-horned profiles (but see \citealt{Schombert2006}). Furthermore, this fraction increases to $30\%$ if only edge-on galaxies are considered, suggesting that the shape of these flat objects can be driven by rotation. Indeed, the projected ellipticity of a rotationally supported oblate spheroid with a constant anisotropy parameter is an increasing function of $v/\sigma$ \citep{Binney1978}, so that objects with prevailing ordered motions are expected to be more flattened. This picture is supported by fig.\,8 of \citet{Geha2006}, where it is shown that the small number of dwarf galaxies with low apparent axis ratios always lie above the best-fit baryonic Tully-Fisher relation \citep{McGaugh2000}, i.e., their rotation is higher than the mean for their baryonic masses.

How do all these results compare with ours? The minimum of our $b/a$ distribution occurs at a characteristic mass (luminosity) M$_{*} \approx 2\times10^{9}$ M$_{\odot}$ ($M_{i}\sim-18$). This is roughly the same value as the lower boundary of Schombert's disc sequence, and would thus indicate a limiting mass for thin discs. The baryonic Tully-Fisher relation derived by Geha et al. indicates that galaxies below these luminosities reside in haloes with $V_{c} \lesssim 80$ km\,s$^{-1}$ and this suggests that if their thickness is set by turbulent motions, the associated temperature floor must be high (see fig.\,1 of KWB07). 

In order to check if stellar feedback effects can be effectively responsible of the previous trend, we have derived the intrinsic thickness of the dwarf galaxies recently modelled by \citet{Governato2010}. These hydrodynamical simulations have first produced realistic dwarf galaxies thanks to their high resolution and a detailed treatment of baryonic processes.  Their simulated DG1 has a stellar mass M$_{*} = 4.8\times10^{8}$ M$_{\odot}$ and a dimensionless spin parameter $\lambda = 0.05$. We have measured a 25 mag\,arcsec$^{-2}$ isophotal axis ratio of $q_{0} = 0.3$ using an $r$-band edge-on image of the dwarf (five-pointed star in Fig.\,1, left). This value is higher than the minimum thickness we obtain for similar mass galaxies, and we speculate as to whether these flatter objects could reside in haloes with higher spin parameters. Interestingly, the second galaxy simulated by Governato et al. (DG2) --which is slightly less massive, has the same $\lambda$ and had a quieter merging history than DG1--, is even thicker ($q_{0} = 0.35$, see Fig.\,1), nicely following the trend with mass we find. We therefore suggest that star formation feedback effects --which can remove low angular momentum material and produce bulgeless dwarfs with shallow central dark matter profiles-- are also responsible of the increasing thickness of low mass galaxies. The shallower potential well certainly allows more turbulent (and therefore also vertical) motions, so stars naturally settle in a thicker disc. 
This scenario is additionally appealing as it can provide an explanation (cf. KWB07) for the observed high gas mass fractions and low star formation efficiencies of faint systems \citep{Warren2007,Amorin2009}.

The results presented in this study have several further implications:
\noindent
\begin{itemize}
\item[1)] The existence of a minimum stellar mass for thin discs has a pronounced impact on the formation of bars, as these structures are thought to be formed through instabilities in cold (thin), rotationally supported discs. Therefore, bar formation could be inhibited in galaxies less massive than M$_{*} \approx 2\times10^{9}$ M$_{\odot}$ simply because they are not cold enough. This is consistent with the result recently obtained by \citet{MendezAbreu2010}, which show that bars are only found in the narrow $10^{9}$-$10^{11}$ M$_{\odot}$ mass range.

\item[2)] It has long been noticed (e.g., \citealt{Bottinelli1983}) that a varying intrinsic thickness can introduce serious systematic errors in calculated inclinations if they are computed assuming discs have a universal, fixed $q_{0}$ --as it is actually usually done when deprojecting H{\sc i} line profiles. While errors are only of a few per cent if the galaxy is still an oblate spheroid, the situation becomes more critical if dwarfs are actually better represented as triaxial ellipsoids (e.g., \citealt{Sung1998}). Moreover, the recent results of \citet{Roychowdhury2010} indicate that gas discs are \emph{also} thicker in faint galaxies, further increasing the previous concerns.

\item[3)] As already pointed out by KWB07, simulations that investigate the role of environmental effects on the population of dwarf galaxies usually consider a low-mass \emph{thin} disc as their fiducial model. Given that real dwarf galaxies are intrinsically thicker --and have shallow central dark matter profiles--, they are likely more susceptible to dynamical interactions than most simulations predict.
\end{itemize}

Interestingly, the evidence of increasing thickness with decreasing mass can be found in several other samples of field late-type dwarfs in the literature (e.g., \citealt{Eder2000,Geha2006})  as well as for non-nucleated dEs in the Virgo cluster \citep{Lisker2007}, further suggesting that it might be related to a formation process independent of environment. We believe that future and detailed hydrodynamical simulations of dwarf galaxy formation --probing a variety of masses, spin parameters and merging histories-- should be able to reproduce the probability distribution of apparent axis ratios and provide more quantitative explanations of the observed trends.

\section*{Acknowledgements}
The authors are happy to thank Fabio Governato and Chris Brook for sharing the images of their simulations and for very instructive discussions. 
Marc Huertas-Company and Ricardo Amor\'in also made helpful comments.
An anonymous referee provided positive feedback that helped improve the paper.
JMA and JALA are partially funded by the Spanish MICINN under the Consolider-Ingenio 2010 Program grant CSD2006-00070: First  Science with the GTC and by the AYA2007-67965-C03-01 project. This paper uses data from the Sloan Digital Sky Survey (http://www.sdss.org).

\bibliographystyle{mn2e}
\bibliography{/Users/rsanchez/WORK/PAPERS/rsj_references.bib}

\label{lastpage}

\end{document}